\documentclass[%
preprint,
showpacs,
 amsmath,amssymb,
 aps,
prl,
a4paper
]{revtex4-1}

\usepackage{graphicx}
\usepackage{dcolumn}
\usepackage{bm}

\begin{document}

\title{Generating carrier-envelope-phase stabilized few-cycle pulses from a
free-electron laser oscillator}

\author{Ryoichi Hajima$^{1)}$}
 \email{hajima.ryoichi@qst.go.jp}

\author{Ryoji Nagai$^{1)}$}%

\affiliation{%
 1) National Institutes of Quantum and Radiological
 Science and Technology,
 Tokai, Naka, Ibaraki 3191106 Japan
}%

\date{\today}

\begin{abstract}
We propose a scheme to generate carrier-envelope-phase (CEP)
stabilized few-cycle optical pulses from a free-electron
laser oscillator.
The CEP stabilization is realized by continuous injection of CEP-stabilized
seed pulses from an external laser to the FEL oscillator whose
cavity length is perfectly synchronized to the electron bunch repetition.
Such CEP-stabilized few-cycle FEL pulses will be an efficient driver
for exploring high-harmonic generation at energies 1-10~keV for
attosecond and zeptosecond science,
broad-band x-ray optical frequency combs
and coherent control of chemical reaction.
\end{abstract}

\maketitle
Recent progress in ultrafast lasers has realized
generation of ultra-intense optical pulses comprising only
a few field oscillation cycles to open new avenues for 
strong-field and attosecond science \cite{Brabec-RMP, Krausz-RMP}.
Stabilization of carrier-envelope phase (CEP), which is the timing of 
the field oscillations with respect to the pulse peak,
is essential to expand applications of such few-cycle optical pulses
to high-harmonic generation (HHG) \cite{Ishii-NC},
 broadband optical frequency combs \cite{Udem-Nature}
and coherent control of chemical reaction \cite{Kling-PCCP}.

Free-electron laser (FEL) oscillators have been operated 
at a wide range of wavelengths from millimeter to VUV,
but CEP stabilization has never been demonstrated
because the evolution of FEL pulses is initiated
by shot noise, microscopic fluctuation of longitudinal density of electrons.
Seeding external laser pulses to single-pass FELs is becoming
a mature technology for improving shot-to-shot wavelength stability,
bandwidth and coherence of FEL pulses \cite{Zhao, Allaria}, but 
CEP-stabilized few-cycle pulses are difficult to generate
even with CEP-stabilized seed laser pulses because of rapid phase
rotation during exponential growth of FEL pulses.

In the present Letter, we propose a scheme 
of continuous injection of CEP-stabilized seed laser pulses
to generate CEP-stabilized few-cycle pulses from FEL oscillators.

The duration of optical pulses generated in a FEL
oscillator is governed by lasing dynamics through the single-pass gain,
round-trip loss of the cavity, electron bunch length, slippage distance and
cavity length detuning \cite{Dattoli,Piovella-PRE-52}.
The slippage distance is defined as $L_s = \lambda N_u$, 
the product of the FEL wavelength, $\lambda$,
and the number of undulator periods, $N_u$.
Cavity length detuning is introduced in FEL oscillators
to compensate for the effect of laser lethargy, i.e., a group velocity
slower than the vacuum speed of light.
In a FEL oscillator in the strong-slippage regime,
in which the electron bunch is shorter than the slippage distance,
an electron bunch superradiantly emits a few-cycle optical pulse
in the limit of small cavity length detuning \cite{Piovella-PRE-52, Chaix-PRE-1999}.
The generation of a 6-cycle pulse at a wavelength of 10.4~$\mu$m
and a 7-cycle pulse at 8.5~$\mu$m from FEL oscillators
has been reported \cite{FELIX,CLIO} and discussed in the context
of degenerate supermodes \cite{Chaix-PRE-1999}.

FEL lasing in a perfectly synchronized optical cavity
(or zero-detuning length) was demonstrated at 
the Japan Atomic Energy Research Institute (JAERI);
the FEL pulse was characterized to be 2.32 optical cycles
at a wavelength of 23.3~$\mu$m \cite{Nishimori-PRL, Hajima-PRL}.
In experimental and numerical studies, it was found that
the lasing in a perfectly synchronized optical cavity only
occurs in the high-gain and small-loss regime of
short-pulse FEL oscillators and requires a relatively long
rising time to reach saturation \cite{Nishimori-PRL, Hajima-sase, Ottaviani},
which is supported by a superconducting linac.

In our proposal, CEP-stabilized few-cycle pulses are
realized by combining FEL lasing in a perfectly synchronized 
optical cavity and an external seed laser with CEP stabilization.
In the following text, the generation of CEP-stabilized few-cycle FEL pulses
is discussed based on the results of time-dependent one-dimensional 
FEL simulations that employ a FEL code similar to the analysis
of the JAERI-FEL \cite{Hajima-PRL}.
An optical pulse evolved in a FEL oscillator reaches to saturation
in which single-pass gain is balanced by round-trip loss.
Transverse profile and phase front of the FEL pulse are primarily
determined by the eigenmodes of the oscillator; thus one-dimensional 
simulations 
yield a reasonable approximation to reproduce lasing behavior in
FEL oscillators \cite{Bakker-1993, Piovella-PRE-52, Nagai-2002}.
Our previous study revealed that the shot noise of an electron beam
plays a critical role in a perfectly synchronized cavity 
not only in initiating the FEL lasing
but also in sustaining the lasing after saturation \cite{Hajima-NIMA-2001}.
In our simulation, the shot noise is implemented according to a model proposed
by Penman and McNeil \cite{Penman} and
coherent spontaneous emission \cite{CSE} is not included.
We assume a machine design similar to the JAERI-FEL but change
the wavelength to 6~$\mu$m considering potential applications of the FEL to HHG
as listed in Table I.

\begin{table}[p]
\caption{\label{tab:table1} Parameters of the FEL oscillators}
\begin{tabular}{lll}
\toprule
           & JAERI-FEL & This Letter \\ \hline
Electron beam & & \\
energy  (MeV) & 16.5 & 50 \\
bunch charge (pC) & 510 & 100 \\
norm. emittance ($x/y$)  & 40/22 & 12/12 \\
\;\;\;\;\;\;\;\;\;\;\;\;\;\;\; (mm-mrad)  &  &  \\
bunch length $^{(*)}$ (ps) & 5  & 0.4  \\
peak current (A) & 200 & 250 \\
bunch repetition (MHz) & 10 & 10 \\ \hline
undulator & \\
undulator parameter (rms) & 0.7 & 1.25 \\
pitch (cm) & 3.3 & 4.5 \\
number of periods  & 52 & 40 \\ \hline
FEL & \\
wavelength  ($\mu$m) & 22.3 & 6  \\
Rayleigh length (m) & 1.0 & 0.52 \\
FEL parameter, $\rho$ & 0.0044 & 0.0052 \\
cavity loss & 6\% & 4\% \\ \hline
\end{tabular}
\par $^{(*)}$ The bunch length is FWHM of 
a triangular bunch for JAERI-FEL and
full width of a rectangular bunch
for the simulations in This Letter.
\end{table}

\begin{figure}[p]
\center
\includegraphics[width=0.9\textwidth]{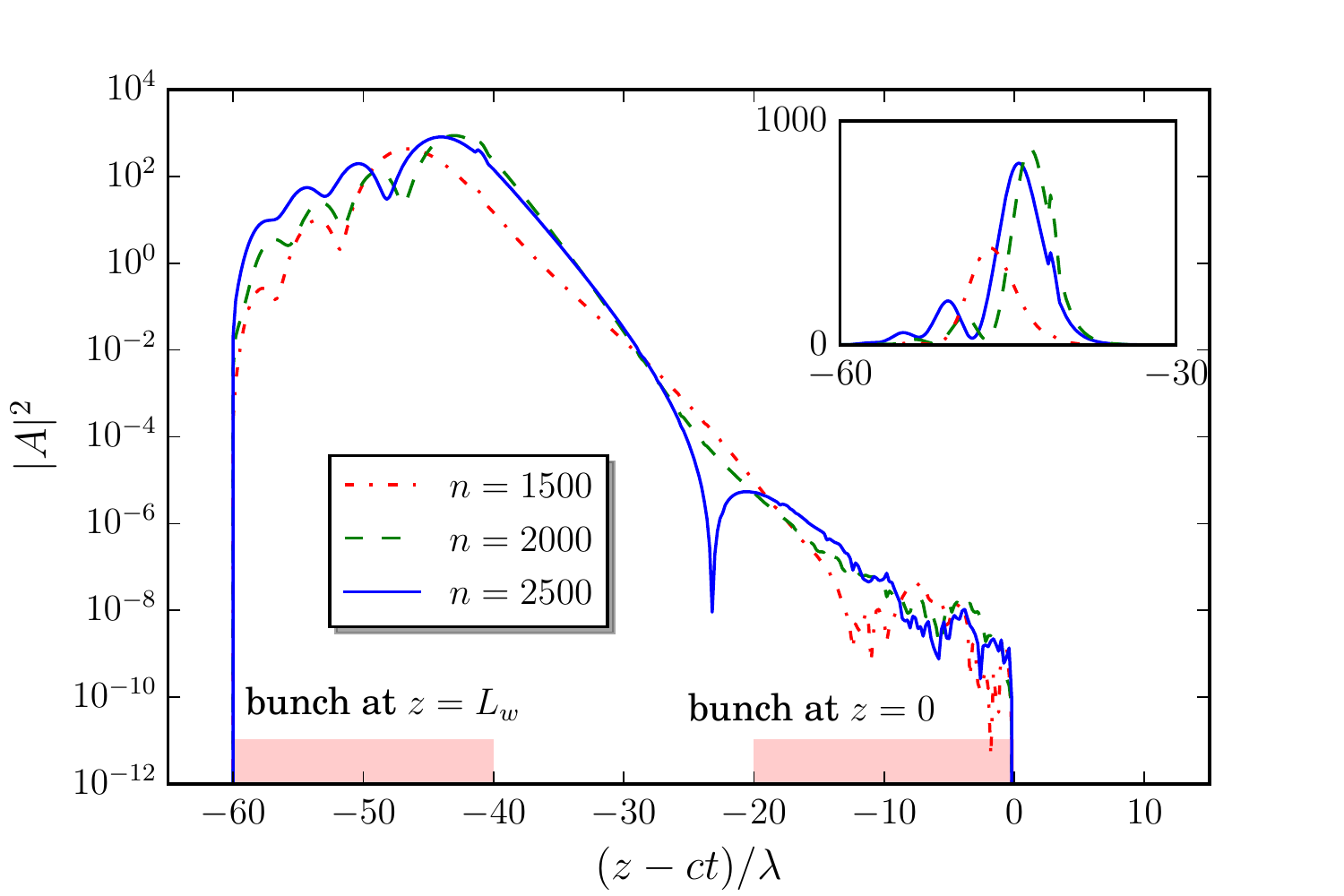}%
\caption{\label{fig:FEL-pulse-random-noise}
 (Color) Temporal shapes of FEL pulses in
a perfectly synchronized optical cavity.
Profile of the electron bunch at the entrance, $z=0$, and
the exit, $z=L_w$, of the undulator is also plotted.
The inset is the same FEL pulses plotted with a linear scale.}
\end{figure}

Figure 1 shows
the temporal profiles of FEL pulses after 1500, 2000 and 2500 round trips
in a perfectly synchronized FEL oscillator.
In this plot, the longitudinal coordinate is defined such that
the leading edge of the electron bunch is located at the undulator entrance,
$z=0$, at the reference time, $t=0$.
The pulse intensity is expressed as a dimensionless value
normalized by the high-gain FEL parameter $\rho$
\cite{Bonifacio}.
The position and profile of the electron bunch at the entrance
and exit of the undulator are also depicted to demonstrate that the FEL lasing
is in the strong-slippage regime.

Figure 1 illustrates characteristics of FEL pulses
evolved in a high-gain and strong-slippage FEL oscillator
with a perfectly synchronized optical cavity.
The optical pulse consists of an exponential lobe of leading edge
and a main peak followed by ringing.
The duration of the main peak,
4.4 cycles (FWHM) after 2500 round trips,
is much shorter than that of the electron bunch.
The pulse height and peak position are not fixed and exhibit continuous
variation along with the pulse energy changes.
A main peak followed by ringing is common to superradiance
observed in two-level systems \cite{Burnham-PR}
and identical to previous results for 
the analysis of a high-gain FEL amplifier \cite{Bonifacio} and 
a perfectly synchronized FEL oscillator 
in the transient regime \cite{Piovella-PRE-51}, both of which
indicated the FEL lasing to be superradiance.
The electron bunch slips backward inside the optical pulse
during the motion in the undulator.
In this motion, the electron bunch forms microbunch through interaction with
the optical field and then emits strong radiation in the slippage region.
The emission in the slippage region is accompanied by frequency down chirp
to keep emission along electron energy decreasing \cite{Hajima-PRL}.
This down chirp contributes to a large FEL conversion efficiency, 
but introduces a large energy spread in the spent electron beam as well.
In a conventional FEL oscillator, the optical cavity length
is shortened so that the optical pulse is pushed forward
every round trip to enlarge a single-pass gain
by introducing strong electron bunching
with a high-intensity optical field at an early section
of the undulator.
The cavity-length shortening achieves feedback of radiation power
from the tail to the head of optical pulse, which also communicate
optical phase and frequency from the tail to the head.
As a result, the cavity length shortening constrains phase correlation 
inside the optical pulse, which prohibits the strong
down chirp observed in the lasing at a perfectly synchronized
cavity.

The logarithmic plot of the FEL pulse in Fig. 1 shows that
the dynamic range of laser pulse intensity from the leading edge to the peak
is greater than $10^{11}$.
The leading part of the optical pulse contains incoherent shot noise
with random amplitude and phase.
The amplitude and phase of the field in the exponential 
envelope, the main peak and the ringing are all governed by
the interaction of electrons and the radiation initiated by the shot noise
in the leading part.
Consequently, the carrier frequency and phase of the FEL pulses are
not stabilized and vary over many round trips.

The simulation result shown in Fig. 1 suggests the possible stabilization
of the optical pulse frequency and phase by fixing the amplitude and phase
of the shot noise in the leading part of the FEL pulse.
This stabilization is realized by overlapping the pulse head with
an external seed laser pulse whose frequency and phase are stabilized.
We conducted a simulation to confirm the scheme of CEP stabilization.
Figure 2 shows FEL pulses obtained 
in a simulated FEL oscillator with injection seeding, 
where all the parameters are the same as in Fig. 1.
The seed laser pulse is assumed to have the resonant wavelength,
an intra-cavity intensity of
$|A_{seed}|^2 = 1.3 \times 10^{-5}$ and a temporal duration of 
$20 \lambda$ with CEP stabilization.
The seed pulse timing is chosen such that half of the seed pulse
overlaps with the FEL pulse and the rest is out of the FEL pulse
to indicate the seed laser intensity not affected by the FEL interaction.

\begin{figure}[p]
\center
\includegraphics[width=0.9\textwidth]{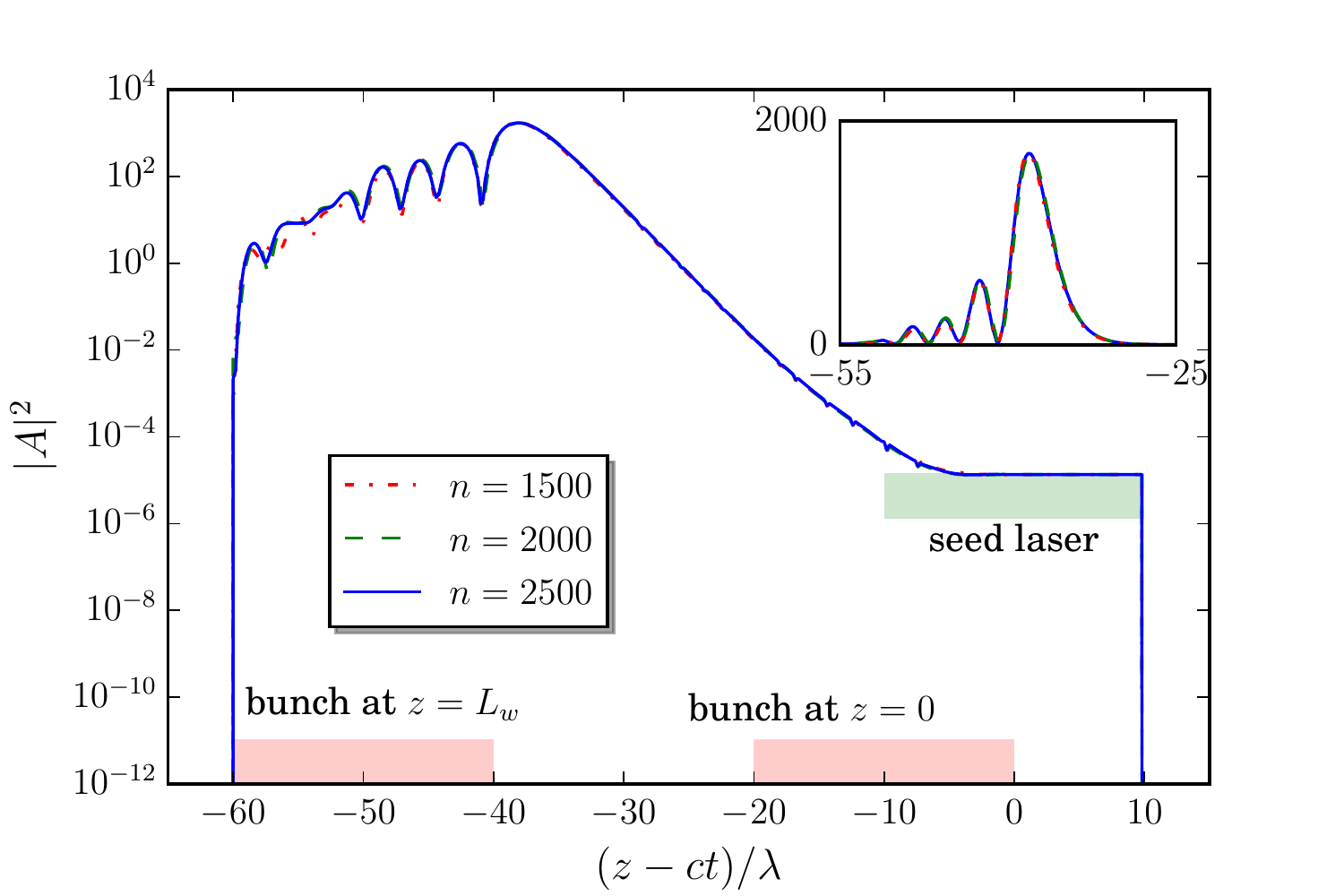}%
\caption{\label{fig:FEL-pulse-with-seed}
 (Color) Temporal shapes of FEL pulses in
a perfectly synchronized optical cavity with
an external seed laser after 
1500, 2000 and 2500 round trips.}
\end{figure}

In Fig.2, we can see that the seed laser 
efficiently stabilizes the FEL oscillator 
with a perfectly synchronized cavity. 
The FEL pulse after the saturation
retains an almost identical temporal shape:
the main pulse of 3.8 cycles (FWHM) followed by periodic ringing. 

\begin{figure}[p]
\center
\includegraphics[width=0.94\textwidth]{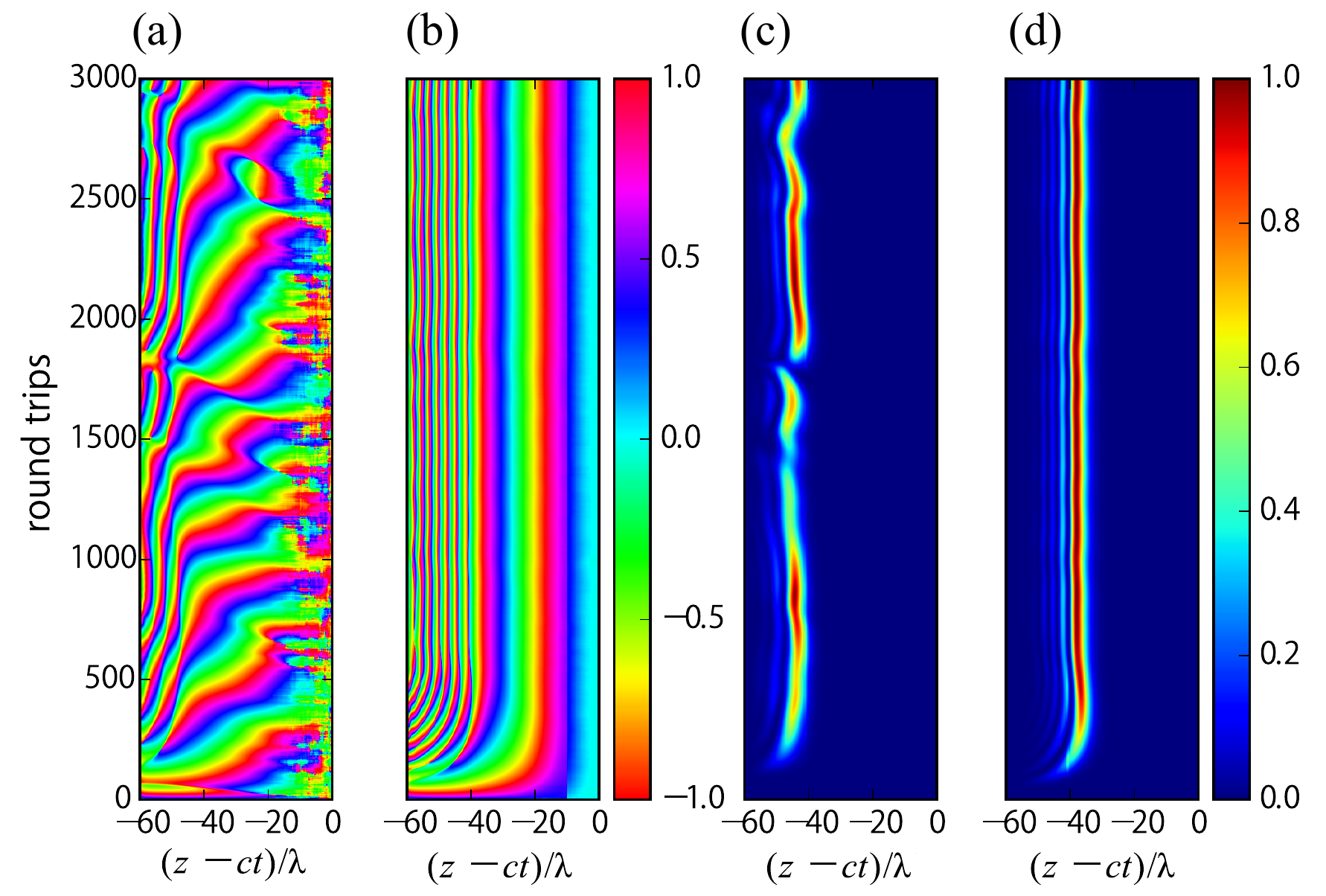}%
\caption{\label{fig:contor}
(Color) Contour plots of instantaneous phase
of FEL pulses in units of $\pi$ rad for 
(a) without injection seeding and (b) with injection seeding.
Contour plots of instantaneous intensity
of FEL pulses normalized to the maximum intensity for
(c) without injection seeding and (d) with injection seeding.}
\end{figure}

The effects of the CEP stabilization can be clearly seen in 
Fig. \ref{fig:contor}, which shows instantaneous intensity
and phase of FEL pulses evolving in a perfectly synchronized cavity
for the two cases without and with a seed laser.
The instantaneous phase, $\phi_L$, is defined such that the complex field
is expressed as $ |A| \exp i (\omega_r (z/c - t) + \phi_L)$,
where $\omega_r$ is the FEL resonance frequency.
The simulation parameters are same as Figs. \ref{fig:FEL-pulse-random-noise}
and \ref{fig:FEL-pulse-with-seed}, respectively.
In the lasing without a seed laser, the phase of the leading part
has fluctuation due to the shot noise,
which makes pulse shape and CEP unstable over many round trips.
The FEL pulse evolution with an external seed laser exhibits
a quite different aspect, in which the pulse shape and CEP after
the onset of saturation are stabilized.
The nonlinear phase advance from the head to the tail of pulses
in Fig. 3 (a) and (b) corresponds to the frequency down chirp \cite{Hajima-PRL}.

For the CEP stabilization,
the seed laser intensity must be sufficiently large compared to
the shot noise intensity.
A set of simulations were performed to determine
the amount of CEP fluctuations and pulse energy fluctuation
as a function of the seed laser intensity.
We plot the simulation result, i.e. 
variation of CEP and pulse energy after saturation
against the intra-cavity seed laser intensity in Fig. 4.
The average intensity of the shot noise at the leading edge of the pulse,
$-20 < (z-ct)/\lambda < 0$, for the simulation parameters
is found to be $|A_{noise}|^2 = 1.83 \times 10^{-8}$, as indicated
by the broken line.
The CEP is uniformly random when the seed laser
intensity is less than the shot noise.
However, the CEP is stabilized for a seed intensity exceeding 
the shot noise level
and the rms error of CEP, $\Delta \phi$, monotonically decreases
with the scale of $ \Delta \phi \propto (|A_{seed}|^2)^{-0.56}$,
which is almost consistent with a phase error equal to the vector sum of 
the seed laser and the random shot noise:
$\Delta \phi_{sum} \propto (|A_{seed}|^2)^{-0.5}$.
The injection seeding also stabilizes the FEL pulse energy.
The FEL conversion efficiency with injection seeding was found
to be 10\%, which corresponds to an intra-cavity pulse energy
of 13~mJ and an extracted pulse energy of 0.5~mJ
neglecting diffraction and absorption losses in the optical cavity.

The fluctuations in the CEP caused by FEL gain variations were also
numerically evaluated for $|A_{seed}|^2 = 1.3 \times 10^{-5}$.
In the simulations, we introduced random jitter into the electron peak
current to vary the FEL gain within a macro pulse
and evaluated the fluctuations in the CEP.
Figure 5 plots the calculated CEP errors as a function of
the amount of jitter in the peak current;
the results indicate that the rms error of CEP
becomes 0.15 rad for a peak current jitter of 20\%.
In a practical design of FELs, accelerator parameters can be optimized
to minimize the jitter, energy spread and emittance at a FEL 
undulator \cite{Wang-JMEA}. 

\begin{figure}[p]
\center
\includegraphics[width=0.9\textwidth]{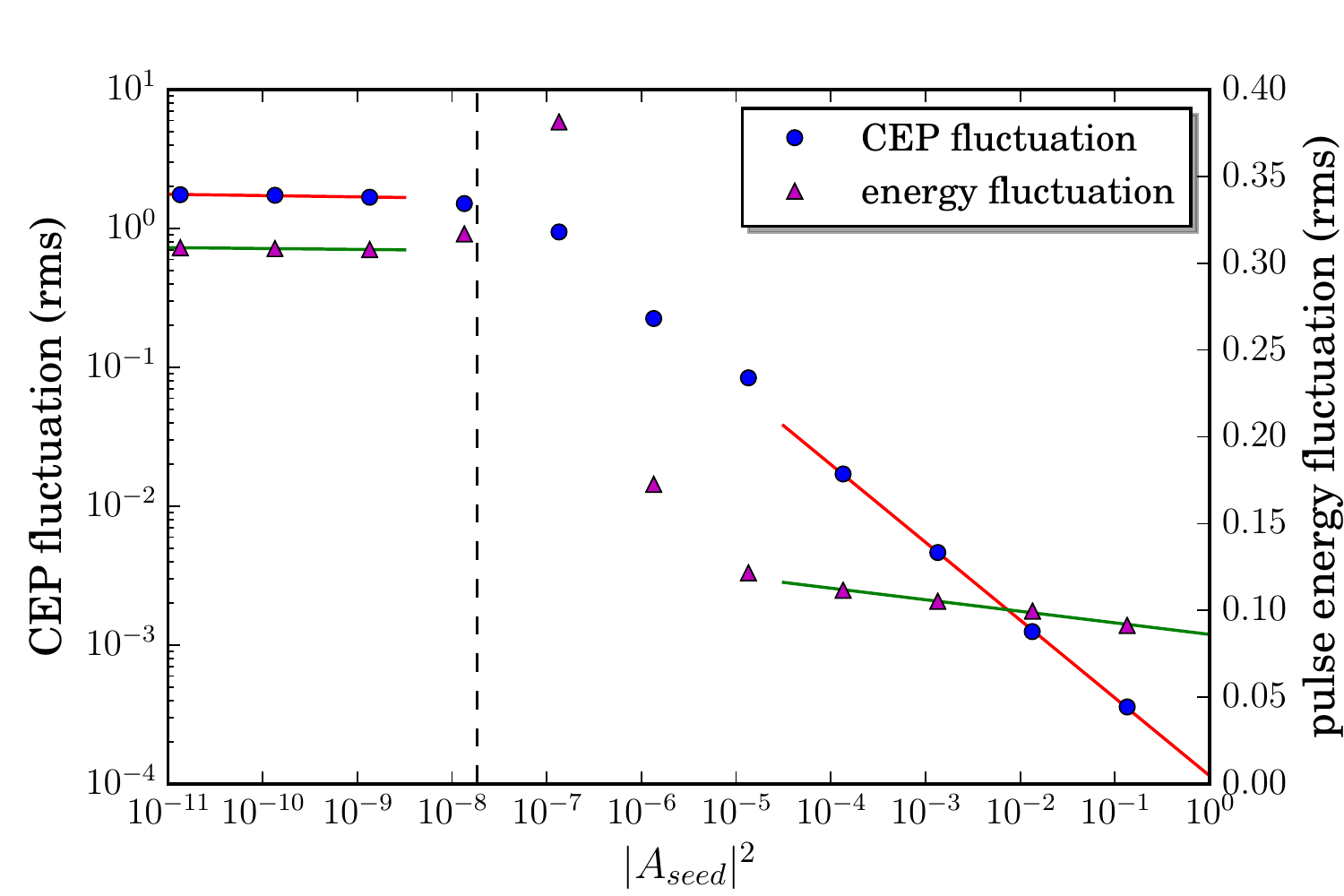}%
\caption{\label{fig:phase-error}
(Color) Fluctuation of carrier-envelope phase 
and pulse energy as a function of
intra-cavity seed laser intensity. The broken line is shot noise intensity.}
\end{figure}

\begin{figure}[p]
\center
\includegraphics[width=0.9\textwidth]{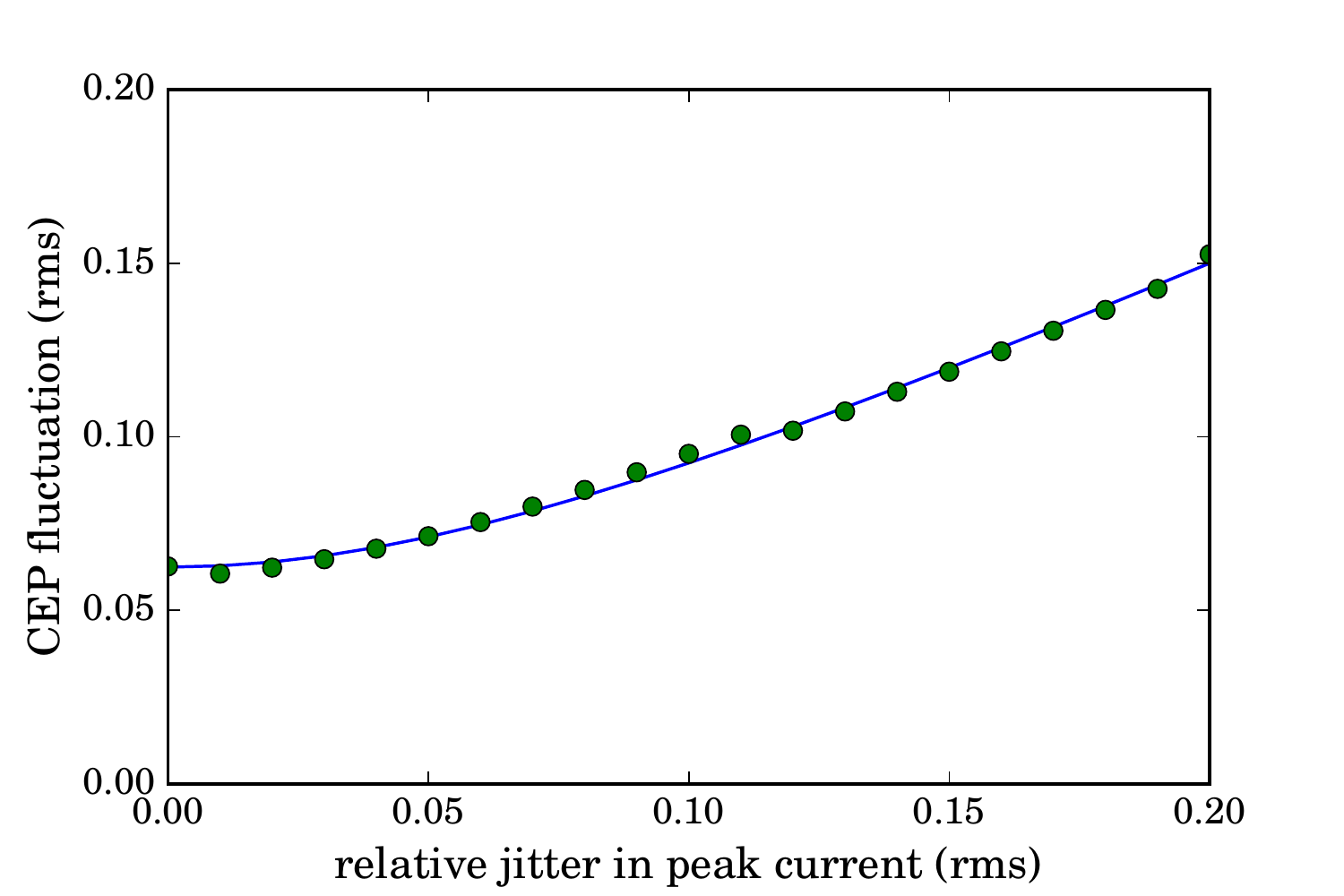}%
\caption{\label{fig:jitter-peak-current}
(Color) Fluctuation of carrier-envelope phase, $\Delta \phi$,
as a function of the rms jitter in the peak current, $\Delta I$.
The solid curve is the best-fit curve employing a function form
$\Delta \phi = \sqrt{a^2 + b^2(\Delta I)^2}$.}
\end{figure}

We comment on the implementation of a seed laser 
for the proposed scheme.
A seed laser must generate a train of CEP-stabilized optical pulses 
at a repetition rate matching that of a train of electron bunches.
The role of the seed laser is inducing an electron energy modulation with 
a stabilized wavelength, amplitude and phase.
The interaction of the electron bunch and seed pulse
can be conducted with various configurations.
The simplest case involves overlapping a seed laser and an electron bunch
in an early section of the FEL undulator.
We can also use two seed lasers with different wavelengths such that
the sum-frequency or difference-frequency of the two lasers is equal to the
FEL wavelength.
Alternatively, a dedicated seeding undulator
installed upstream of the FEL undulator also works for CEP stabilization.
In the dedicated undulator, we can 
control polarization of the seed laser and the FEL independently
and use a sub-harmonic of the FEL as a seed laser wavelength.

In the above simulations, we assumed a single-undulator configuration,
in which the seeding and the FEL lasing share the undulator 
and the optical cavity.
The FEL cavity length should be controlled such that the round-trip time
of the FEL pulse matches the electron bunch interval.
Moreover, the cavity length must satisfy the condition for
intra-cavity coherent stacking of successive seed pulses.
Such control has been established for laser Compton scattering experiments
employing a laser enhancement cavity \cite{Akagi}.
The intensity of the seed pulse in Fig. 2,
$|A_{seed}|^2 = 1.3 \times 10^{-5}$,
corresponds to an intra-cavity pulse energy of 0.34 nJ for a 400-fs pulse.

In conclusion, we have proposed a scheme to generate CEP-stabilized few-cycle
laser pulses in a FEL oscillator with a CEP-stabilized external seed
laser. The proposed scheme is applicable to a wide range of wavelengths
from infrared to terahertz as far as a high-gain strong-slippage
FEL oscillator and a CEP-stabilized seed laser are available.
Such FEL pulses can be delivered to either extra-cavity or intra-cavity
experiments \cite{FELICE}.
Few-cycle mid-infrared FEL pulses with mJ-class energy
are attractive because they can enable HHG,
thereby producing vacuum ultraviolet and X-ray pulses
 with attosecond duration \cite{Tecimer}.
Because the HHG cutoff energy under the phase-matched condition
depends on the drive laser wavelength 
as $h \nu _{cutoff} \propto \lambda^{1.7}$,
many efforts are focused on extending the drive laser wavelength 
from the near-infrared
to the mid-infrared \cite{Popmintchev-Science,Cavaletto-Nat-Phot}.
The few-cycle mid-infrared FEL pulses at 10-MHz repetition proposed here
could be a unique method for exploring HHG beams at energies of 1-10~keV
to push ultrafast laser science to the zeptosecond regime
and for realizing X-ray optical frequency combs.

The authors thank Xiao-Ming Tong and Jiro Itatani for discussion
regarding HHG and optical frequency combs.
This research was partially supported by 
the Research Foundation for Opto-Science and Technology.

\end{document}